\documentstyle[12pt]{article}
\begin{document}

\begin{center}
{\bf GAUGE INVARIANCE FOR THE MASSIVE AXION}
\end{center}

\begin{center}
{\bf  P.J.Arias$^{a,b,}$\footnote{email: parias@tierra.ciens.ucv.ve}
and A.Khoudeir$^{b,}$\footnote{adel@ciens.ula.ve}}\\
  
${}^a${\it Grupo de Campos y Part\'{\i}culas, Departamento de
F\'{\i}sica, 
Facultad de \\ 
Ciencias, Universidad Central de Venezuela, AP 47270, Caracas 1041-A,
Venezuela.}\\[4mm]
${}^b${\it Centro de Astrof\'{\i}sica Te\'orica, Facultad de Ciencias,
 Universidad de los Andes, La Hechicera, M\'erida 5101, Venezuela.}\\
[15mm]
\end{center}
\begin{center}
{\bf ABSTRACT}
\end{center}

A massive gauge invariant formulation for scalar ($\phi$) and 
antisymmetric ($C_{mnp}$) fields with a topological coupling, 
which provides a mass for the axion field, is considered. 
The dual and local equivalence with the non-gauge 
invariant proposal is established, but on manifolds with non-trivial 
topological structure both formulations are not globally equivalent. 
\newpage
\section{INTRODUCTION}

In four dimensions a massless (pseudo)scalar field: the axion, is 
dual to the antisymmetric field $B_{mn}$ (only if derivative 
couplings are considered, therefore massive terms are excluded) as a 
particular case of the general duality between $p$ and $D-p-2$ forms 
in $D$ dimensions. Since non-perturbative effects break the local 
Peccei-Quinn symmetry for the axion field, in order to give mass 
to the axion, the duality between a massive axion and an 
antisymmetric field was considered an enigma until two independent 
approaches \cite{linde}, \cite{quev} were developed recently. 
Now, is understood that take into account non-perturbative effects, 
the usual duality between a massless scalar field and 
an antisymmetric field $B_{mn}$ is not broken, but 
replaced by the duality between a massive scalar $\phi$ and an 
massive antisymmetric field $C_{mnp}$. An early attempt to undestand 
this duality was considered in the reference\cite{curt}
A characteristic feature of this duality is the lost of abelian gauge 
invariance for the antisymmetric field. In this article, we will show 
that a gauge invariant theory, which involved a topological coupling and 
considered several years ago \cite{aur} in the context of the $U(1)$
problem, 
is locally equivalent to the non gauge invariant proposal. This 
equivalence is similar to what happen in three dimensions for 
massive topologically and self-dual theories \cite{djvn} and the 
Proca and massive topologically gauge invariant theories in four 
dimensions \cite{pio}. We will study the equivalence through the 
existence of a master action from which local and global 
considerations are established.

\section{THE GAUGE INVARIANT MODEL}

An illustrative model for the massive axion is given by the following 
master action\cite{quev1}
\begin{equation}
I = <-\frac{1}{2}v_m v^m + \phi\partial_{m}v^{m} + \frac{1}{2}m^2 \phi^2
>,
\end{equation}
where $v_m$ is a vector field and $\phi$ is a scalar field($< >$ 
denotes integration in four dimensions).
Eliminating the field $v_m $ through its equation of motion: $v_m = 
\partial_m\phi$, the action for a massive scalar is obtained, while
using 
the equation of motion, obtained by varying the scalar field $\phi$ 
($\phi = \frac{1}{m^2}\partial_m v^m$), we have 
\begin{equation}
I_v = \frac{1}{2}< v_m v^m + \frac{1}{m^2}(\partial_m v^m)^2 >
\end{equation}
and the propagator corresponding to the field $v_m$ is 
$\eta_{mn} - \frac{k_{m}k_{n}}{k^2 + m^2}$, which is just equal to those 
discussed in \cite{linde}.   
A simple way to show the duality, rely on introducing 
the dual of the vector field $v^m = \frac{1}{3!}m\epsilon^{mnpq}C_{npq}$ 
in the action $I_v$, yielding the master action
\begin{equation}
I_{M1} = < -\frac{1}{2.3!}m^2 C^{mnp}C_{mnp} +
\frac{1}{3!}m\epsilon^{mnpq}
\phi\partial_{m}C_{npq} - \frac{1}{2}m^2 \phi^2 >.
\end{equation}
from which the duality is easily infered.
In fact, eliminating the scalar field $C_{mnp}$ (or $\phi$) through its 
equation of motion, the action for a massive scalar field (or the
massive 
antisymmetric field $C_{mnp}$) is obtained. In anycase, the gauge
invariance is 
spoiled. Now, we can ask whether there really exist an invariant gauge 
theory compatible with a massive term for the axion field. 
The answer is possitive. We will show that the following action
\begin{equation}
I_{M2} = <-\frac{1}{2}\partial_m\phi\partial^m\phi -
\frac{1}{2.4!}G_{mnpq}
G^{mnpq} - \frac{m}{6}\epsilon^{mnpq}C_{mnp}\partial_q\phi >,
\end{equation}
where $G_{mnpq} \equiv \partial_m C_{npq} - \partial_n C_{mpq} + 
\partial_p C_{mnq} - \partial_q C_{mnp}$ is the field strenght
associated 
to the antisymmetric field $C_{mnp}$, is locally equivalent to $I_{M1}$, 
describing the propagation of a massive scalar excitation: a massive
axion. 
Note that the coupling term is an extension of the usual $BF$ term and 
the action is invariant under the abelian gauge transformations
\begin{equation}
\delta_{\xi} C_{mnp} = \partial_{m}\xi_{np} + \partial_{n}\xi_{pm} + 
\partial_{p}\xi_{mn}, \quad \delta_{\xi}\phi = 0.
\end{equation}
This action was considered previously in ref \cite{aur}. as a
generalization 
to four dimensions of the Schwinger model in two dimensions.
 
Let us see, how this action is related to 
the propagation of a massive axion and why the equivalence with the 
non gauge invariant action must hold.
Rewritten down the action (eq. (4))
by introducing $F^{mnp} \equiv \epsilon^{mnpq}F_q $ as 
the dual tensor of $F_m = \partial_m \phi$, we can eliminate 
$F^{mnp}$ through its equation of motion: $F^{mnp} = -mC^{mnp}$ and 
substituing, the action for the massive antisymmetric field 
$C^{mnp}$ appears. Going on an additional step, the dual of 
the antisymmetric field $C^{mnp} = \frac{1}{m}\epsilon^{mnpq}v_q $ 
is introduced, and the action for the vector field $v_m$, eq. (2), is
obtained. 
On the other hand, if we introduce $\lambda \equiv
-\frac{1}{4}\epsilon^{mnpq}
G_{mnpq}$ as the dual of the strenght field $G_{mnpq}$ into the action
(4), 
we observe that $\lambda$ plays the role of an auxiliary field, whose 
elimination through its equation of motion ($\lambda = -m\phi$) lead to 
the action of a massive scalar field. 

It is worth recalling, since the action is expressed only in derivatives 
of the scalar field, that the dual theory can be achieved, 
reemplacing $\partial_m \phi$ by $\frac{1}{2}l_{m}$ and add a BF term: 
$\frac{1}{4}l_m\epsilon^{mnpq}\partial_n B_{pq}$\cite{nt}. The dual
action 
is\cite{aur}
\begin{equation}
I_d = < -\frac{1}{2.4!}G_{mnpq}G^{mnpq} - \frac{1}{2.3!}
(mC_{mnp} - H_{mnp})(mC^{mnp} - H^{mnp}) >,
\end{equation}
where $H_{mnp} = \partial_m B_{np} + \partial_n B_{pm} + \partial_p
B_{mn}$ is 
the field strength of the antisymmetric field $B_{mn}$, which was
introduced 
in the BF term. This action just describes the interaction of open
menbranes 
whose boundaries are closed strings \cite{at} and is invariant under the 
following gauge transformations
\begin{equation}
\delta C_{mnp} = \partial_{m}\xi_{np} + \partial_{n}\xi_{pm} + 
\partial_{p}\xi_{mn}, \quad \delta B_{mn} = \partial_m\lambda_n - 
\partial_n\lambda_m - m\xi_{mn}.
\end{equation}
The $\xi$ gauge transformation allows us gauged away the antisymmetric
field 
$B_{mn}$, leading to the massive antisymmetric field $C_{mnp}$ action.

\section{THE EQUIVALENCE}

Now, we are going on to show the equivalence. Let us take the following 
master action
\begin{equation}
I_M = <-\frac{1}{3!}m^2 a_{mnp}a^{mnp} - \frac{1}{2!}m^2\psi ^2 + 
\frac{1}{4!}m\epsilon^{mnpq}\psi G_{mnpq} + \frac{1}{3!}m\epsilon^{mnpq}
(a_{mnp} - C_{mnp})\partial_q\phi>.
\end{equation}
Independent variations in $a_{mnp}, \psi, C_{mnp}$ and $\phi$ lead to
the 
following equations of motion
\begin{equation}
a^{mnp} = \frac{1}{m}\epsilon^{mnpq}\partial_q\phi,
\end{equation}
\begin{equation}
\psi = \frac{1}{4!m}\epsilon^{mnpq}G_{mnpq},
\end{equation}
\begin{equation}
\epsilon^{mnpq}\partial_{m}(\psi - \phi) = 0,
\end{equation}
and 
\begin{equation}
\epsilon^{mnpq}\partial_{q}(a_{mnp} - C_{mnp}) = 0.
\end{equation}
Replacing the expressions for $a_{mnp}$ and $\psi$ given by eqs. (9) and
(10) 
into $I_M$, the gauge invariant action $I_{M2}$ is obtained. 
On the other hand, the solutions of the equations of motion 
(11) and (12) are
\begin{equation}
\phi - \psi = \omega, \quad C_{mnp} - a_{mnp} = \Omega_{mnp},
\end{equation}
where $\omega$ and $\Omega_{mnp}$ are $0$ and $3$-closed forms,
respectively. 
Locally, we can set 
\begin{equation}
\omega = constant, \quad \Omega_{mnp} \equiv L_{mnp} = \partial_m l_{np}
+ 
\partial_n l_{pm} + \partial_p l_{mn},
\end{equation}
and subtituing into $I_M$, we obtain the following "Stuckelberg" action
\begin{equation}
I_s = <-\frac{1}{3!}m^2 (C_{mnp} - L_{mnp})(C^{mnp} - L^{mnp}) -
\frac{1}{2}
m^2 (\phi - \omega)^2 + \frac{1}{4!}m\epsilon^{mnpq}(\phi -
\omega)G_{mnpq}>.
\end{equation}
This action is invariant under 
\begin{equation}
\delta_{\xi} C_{mnp} = \partial_{m}\xi_{np} + \partial_{n}\xi_{pm} + 
\partial_{p}\xi_{mn}, \quad \delta_{\xi}l_{mn} = \xi_{mn},
\end{equation}
which allow us gauged away the $l_{mn}$ field and recover $I_{M1}$ (we 
have redefined $\phi -\omega$ as $\phi$ since $\omega$ is a constant).
In 
this way, the local equivalence is stated. 
This local equivalence can also be established from a hamiltonian 
point of view and will be reported elsewhere \cite{khar}.
In an ample sense, we 
must consider $\psi = \phi - \omega$ and $a_{mnp} = C_{mnp} -
\Omega_{mnp}$ 
as the general solutions and $I_{M2}$ is locally and globaly equivalent 
to 
\begin{eqnarray}
{\bar I}_{M1} &=& <-\frac{1}{3!}m^2 (C_{mnp} - \Omega_{mnp})(C^{mnp} - 
\Omega^{mnp}) - 
\frac{1}{2!}m^2(\phi - \omega) ^2 \\ \nonumber
&+& \frac{1}{4!}m\epsilon^{mnpq}(\phi - \omega) G_{mnpq} - 
\frac{1}{3!}m\epsilon^{mnpq}\Omega_{mnp}\partial_q\phi>,
\end{eqnarray}
which is an adequate extension of $I_{M1}$. 

Finally, we can eliminate $\phi$ and $C_{mnp}$ to achieve
\begin{equation}
{\bar I}_{M1} = I_{M1[a,\psi]} - I_{top[\omega ,\Omega]},
\end{equation}
where
\begin{equation}
I_{top[\omega ,\Omega]} = < \frac{1}{3!}m\epsilon^{mnpq}\Omega_{mnp}
\partial_{q}\omega >
\end{equation}
is the extension of the BF term for the topological coupling between 
$0$ and $3$-forms in four dimensions. From this result, we have that 
the partition funtions of $I_{M1}$ and $I_{M2}$ differ by a topological
factor.
\begin{equation}
Z_{M2} = Z_{top}Z_{M1}
\end{equation}
In general, on manifolds with non trivial topological structure 
$Z_{top} \neq 1$. Only when the manifold has a trivial structure, 
we must have $Z_{top} \equiv 1$, reflecting the local equivalence.

Sumarizing, we have seen that a gauge invariant description for 
massive axions is possible which is (locally)equivalent to the 
non-gauge invariant proposal. Several aspects of this proposal 
are under considerations: a detailed hamiltonian description for 
both proposal of generating mass for the axion and 
a complete BRST analysis of the gauge invariant model considered 
in this paper\cite{khar}.

\end{document}